# ANALOGICAL MODELING AND QUANTUM COMPUTING

Royal Skousen
Professor of English and Linguistics
College of Humanities
3187 Jesse Knight Humanities Building
Brigham Young University
Provo, Utah 84602

royal_skousen@byu.edu

**Abstract**

This paper serves as a bridge between quantum computing and analogical modeling (a general theory for predicting categories of behavior in varying contexts). Since its formulation in the early 1980s, analogical modeling has been successfully applied to a variety of problems in language. Several striking similarities between quantum mechanics and analogical modeling have recently been noted: (1) traditional statistics can be derived from a non-statistical basis by assuming data occurrences are accessed through a spin-up state (given two equally probable quantum states, spin-up and spin-down); (2) the probability of predicting a particular outcome is determined by the squaring of an underlying linear measure and is the result of decoherence (which occurs when a quantum system is observed); and (3) a natural measure of certainty (called the agreement) is based on one chance of guessing the right outcome and corresponds to the integrated squaring of Schrödinger's wave equation.

Analogical modeling considers all possible combinations of a given context of $n$ variables, which in classical terms leads to an exponential explosion on the order of $2^n$. This paper proposes a quantum computational solution to this exponentiality by applying a cycle of reversible quantum operators to all $2^n$ possibilities, thus reducing the time and space of analogical modeling to a polynomial order.

# Preface

During the last two decades, as rule approaches have encountered difficulties in describing language behavior, several competing non-rule approaches to language have been developed. First was the development (or rejuvenation) of neural networks, more commonly known in linguistics as connectionism and best exemplified by the work of McClelland, Rumelhart, and the PDP Research Group (1986) in what they call "parallel distributed processing" (PDP). More recently, some researchers (such as David Aha and Walter Daelemans) have turned to exemplar-based systems (sometimes known as instance-based systems or "lazy learning") to describe language behavior (Aha, Kibler, and Albert 1991; Daelemans, Gillis, and Durieux 1994). These exemplar-based learning systems involve hunting for the most similar instances ("nearest neighbors") to predict language behavior. A more general theory of the exemplar-based approach is Royal Skousen's analogical modeling of language, which permits (under well-defined conditions) even non-neighbors to affect language behavior.

The fundamental works on analogical modeling are two books by Skousen. The first one, *Analogical Modeling of Language* (Skousen 1989), provides a complete, but basic, outline of the approach (chapter 2) and then shows how the theory can be applied to derive various language properties (chapter 3) as well as deal with several theoretical language issues (chapter 4). In chapter 5, Skousen provides an in-depth analysis of past-tense formation in Finnish. In particular, he shows how analogical modeling, unlike traditional rule approaches, is able to describe the complex historical development of the Finnish past tense. The second book, *Analogy and Structure* (Skousen 1992), is a mathematical description of both rule-based and analogical approaches to describing behavior.

Since the publication of Skousen 1989, analogical modeling has been applied to a number of specific language problems. Derwing and Skousen (1994) have used analogical modeling to predict English past-tense formation, especially the kinds of errors found in children's speech. There have also been a number of applications to several non-English language problems, for instance, German plural formation (Wulf 1996), Spanish stress assignment (Eddington 2000b), and Turkish morphology (Rytting 2000).

An important application of analogical modeling is found in Jones 1996. Here analogical modeling is applied to automatic translation (between English and Spanish). Most work done in analogical modeling has dealt with phonology and morphology, but here Jones shows how analogical modeling can be applied to syntax and semantics. He contrasts analogical modeling with both traditional rule approaches and connectionism (parallel distributed processing). In a variety of test cases, he finds analogical modeling more successful and less arbitrary than parallel distributed processing.

For an overview of analogical modeling, see the following article at <http://humanities.byu.edu/aml/homepage.html>:

Royal Skousen, Analogical Modeling

> to appear in *Quantitative Linguistics: An International Handbook*, edited by Gabriel Altmann, Reinhard Köhler, and Raimund G. Piotrowski (Berlin: Walter de Gruyter, 2001)

**The exponential explosion and quantum computing**

Analogical modeling, from the very beginning, has proposed that in predicting behavior all possible combinations of variables should be tested. If there are *n* variables for a given context, there will be $2^n$ supracontexts (or combinations of variables) to consider. Basically, increasing the specification by one variable doubles the memory requirements as well as the running time (section 6.1 of Skousen 1989; also see Daelemans, Gillis, and Durieux 1997). There have been numerous attempts to deal with this intractability: fine-tuning the computer program, revising the algorithm so that it would not have to keep track of every possible supracontext, and using parallel processing.

A new approach to dealing with the problem of the exponential explosion in analogical modeling has been to re-interpret analogical modeling in terms of quantum computing. (For a general introduction to quantum computing, see Williams and Clearwater 1998; Lo, Popescu, and Spiller 1998; Berman, Doolen, Mainieri, and Tsifrinovich 1998; or Hey 1999.) One distinct theoretical advantage of quantum computing is that it can simultaneously keep track of an exponential number of states (such as $2^n$ supracontexts defined by an *n*-variable given context), thus potentially reducing intractable exponential problems to tractable polynomial analyses (or even linear ones). In certain well-defined cases it has been shown (in pseudo-code only, since there is no complete hardware implementation of quantum computing thus far) that the exponential aspects of programming can be reduced to one of polynomial degree (which entails tractability, unlike exponential cases). Quantum computing allows for certain kinds of simultaneity or parallelism that exceeds the ability of normal computing (sequential or parallel). The examples discussed thus far in quantum computing involve numbers, especially cryptography, as in Peter Shor's algorithm for determining the prime factors of a long integer (see, for example, Williams and Clearwater 1998:133-137).

One reason for considering quantum analogical modeling is that the exponential factor seems to be inherent in all approaches to language processing. Thus far, linguistic evidence argues that virtually all possible combinations of variables can be used by native speakers in predicting language. The exponential problem is explicitly required in analogical modeling, and normal kinds of parallel processing will probably fail to solve this problem. Nor is the exponential explosion in predicting language restricted to analogical modeling. Other exemplar-based approaches and neural networks (connectionist approaches) also encounter exponential problems since researchers using these non-declarative approaches must decide how to limit their predictions to those based on the "most significant" variables. The difficulty for these other approaches is in the training stage, where the system has to figure out which combinations of variables are significant, a global task that is inherently exponential.

In the early 1980s, as Skousen was writing *Analogy and Structure* and setting down the basic principles of analogical modeling, he had no idea of its possible connection with quantum mechanics or the possibility that quantum computing might be used to do analogical modeling. Of course, at that time there was only the initial formulation of what quantum computation might involve (as in Feynman's early ideas and Deutsch's universal quantum computer, plus Landauer's and Bennett's earlier work on reversible computation). Skousen's motivation for analogical modeling was linguistic, although in its mathematical formulation in *Analogy and Structure* considerable attention was paid to measures of uncertainty and accounting for the general nature of rule systems.

The original characterization of analogical modeling has surprisingly remained unchanged over the last two decades. Its application to a number of linguistic problems (both general and specific) has shown that analogical modeling continues to make the right predictions, perhaps because of its similarity with quantum mechanics, a theory which has been successfully applied to virtually all aspects of physical behavior since its first formulation in the 1920s. More recently, there has been an important realization that quantum reality and information theory are closely related, emphasized, for instance, by John Archibald Wheeler (see his article "Information, Physics, Quantum: The Search for Links" in Hey 1999:309-336). The close relationship between analogical modeling and information theory implies that the striking similarities between analogical modeling and quantum mechanics may not be accidental at all -- that in actuality the mechanisms used by speakers of languages to learn and use language may involve quantum computing.

One advantage of analogical modeling is that no mathematical (or statistical) calculation is actually used in determining the analogical prediction; instead, there is just the simple comparison of deterministic and non-deterministic supracontexts. This kind of decision-making process is based on what is referred to as a natural statistic. Natural statistics are psychologically plausible and avoid any direct consideration of probability distributions, yet have the ability to predict stochastic behavior as if some underlying probability distribution were known. The simplicity of analogical modeling suggests that some very basic operators could be used to determine a quantum analogical set that would then be reduced to a single supracontext (combination of variables) whenever decoherence (or observation) occurs.

**Similarities between analogical modeling and quantum computing**

One reason for pursuing the possibility of quantum computing of analogical modeling is that a number of striking similarities have been discovered between analogical modeling and quantum mechanics:

(1)   Traditional statistics assume some complicated underlying mathematical functions, but from natural statistics (which involve no direct numerical calculations) we can derive the results of standard statistics if we assume that the probability of remembering any given data occurrence equals precisely one-half. This relationship implies that traditional statistics can be derived from natural statistics if data occurrences are accessed through, say, a spin-up state (given two equally probable quantum states, spin-up and spin-down).

(2) In both quantum mechanics and analogical modeling, there is an underlying linearity as well as an observed squaring. In quantum mechanics, prior to observation, an exponential number of quantum states can be simultaneously accounted for, yet when observed, this superposition of many states is collapsed into a single one, a process referred to as decoherence. Prior to observation, each quantum state is assigned an amplitude, but this amplitude is squared to give a probability when observation occurs. A single observation leads to this decoherence and squaring of the amplitude. In analogical modeling, there is an exponential number of supracontexts (combinations of variables) for a given context. We keep track of the number of occurrences (a linear function) for each supracontext. When we come to predicting an outcome, one of the supracontexts is selected and the probability of selecting that supracontext is proportional to the square of the number of occurrences in that supracontext. The squaring naturally results from selecting a pointer to an occurrence rather than directly selecting an occurrence.

(3) In analogical modeling, a quadratic measure of agreement is used to measure certainty. Agreement is based on the idea that one gets a single chance to determine the outcome. This single observation corresponds to the decoherence that occurs when a quantum system is observed. Moreover, this measure of agreement corresponds to Schrödinger's wave equation, where squaring is used to determine the probability of occurrence.

In the next three sections, these points are discussed in some detail.

**Traditional statistics from natural statistics**

While investigating natural statistics, Skousen (1998) discovered that when the probability of remembering is one-half, we get standard statistical results (including the ability to account for the traditional "level of significance" used in statistical decision making). However, there seemed to be no inherent motivation for why this one-half probability of remembering should lead to traditional statistics. But the one-half probability can be justified if we interpret it as corresponding to storing the individual occurrences of a database by means of a vector composed of quantum bits, each with an equal chance of being accessed or not (much like an electron's spin, with its two states of up and down).

There are two specific results from natural statistics that argue for the significance of the one-half probability of remembering (Skousen 1998:247-250). First, consider the task of estimating the probability of occurrence $p$ for an outcome. Suppose we have two possible outcomes, either $s$ or $t$. Suppose further that we have been given the following string of outcome data:

$$s\;s\;s\;t\;s\;t\;t\;t\;t\;s\;t\;s\;t\;t\;t\;t\;s\;s\;t$$

If we have perfect memory (where the probability of remembering $r$ is one), then in natural statistics, the probability $p$ of predicting the $s$ or $t$ outcome is directly proportional to the relative frequency of each outcome in the data. So in this string of occurrences, where there are 8 examples of $s$ and 12 of $t$, we get the following predictions under perfect memory ($r = 1$):

$$p(s) = 8/20 = 0.4 \qquad p(t) = 12/20 = 0.6$$

When memory is perfect, we always get this same estimated probability $p$ for the outcome $s$ (namely, 0.4); in other words, there is no variance in our estimate for $p$:

$$\mathrm{Var}(p) = 0 \text{ if } r = 1 \text{ (perfect memory)}$$

Suppose there are $n$ occurrences in the data and that $m$ occurrences are remembered. We can first show that the expected value E of the probability $p$ of an outcome is simply the probability of that outcome -- that is, we have an unbiased estimator for $p$:

$$E(p) = p(\text{outcome})$$

When we consider the variance for this estimator, we get the following relationship (Skousen 1998:248):

$$\text{Var}(p) = 1/(n-1) \cdot E(p)(1-E(p)) \cdot (E(n/m)-1)$$

The two expectations, $E(p)$ and $\text{Var}(p)$, hold no matter what $r$, the probability of remembering, is.

When $r = 1/2$, a given data occurrence is remembered -- or is accessible -- half the time (on the average). Under these conditions and for large $n$, the number of remembered occurrences ($m$) is approximately equal to $n/2$. Thus the expected value for the ratio $n/m$ will be approximately equal to 2. This means that for large $n$ we get the following asymptotic relationship for the variance of $p$ when $r = 1/2$:

$$\text{Var}(p) \approx 1/(n-1) \cdot E(p)(1-E(p))$$

Now this asymptotic measure of variance derived from natural statistics (but only when the memory is 1/2) is precisely the same as the traditional unbiased estimate of variance (which assumes that the relative frequency is first used to estimate $p$).

Now consider a second statistical task. Suppose we have some data with the two outcomes $s$ and $t$, and we want to predict the most frequent of these two outcomes. For simplicity of calculation, suppose our outcome data for this example consists of only the following four occurrences:

*s s s t*

Now the chances of the $s$ outcome being more frequent than the $t$ outcome is assured if we have perfect memory (when $r = 1$). Under such conditions, there will always be three occurrences of $s$ and one of $t$, so there will be no uncertainty in our prediction:

$$p(s > t) = 1 \text{ if } r = 1$$

On the other hand, when $r = 1/2$, each occurrence of the four will be remembered -- or accessed -- half the time (on the average), which will thus give 16 equally possible cases:

**16 sets of remembered occurrences**

|        |         | $p(s)$ | $p(t)$ | $p(\emptyset)$ |
|--------|---------|--------|--------|----------------|
| 1/16   | s s s t | 1/16   | -      | -              |
| 1/16   | s s s - | 1/16   | -      | -              |
| 1/16   | s s - t | 1/16   | -      | -              |
| 1/16   | s - s t | 1/16   | -      | -              |
| 1/16   | - s s t | 1/16   | -      | -              |
| 1/16   | s s - - | 1/16   | -      | -              |
| 1/16   | s - s - | 1/16   | -      | -              |
| 1/16   | s - - t | 1/32   | 1/32   | -              |
| 1/16   | - s s - | 1/16   | -      | -              |
| 1/16   | - s - t | 1/32   | 1/32   | -              |
| 1/16   | - - s t | 1/32   | 1/32   | -              |
| 1/16   | s - - - | 1/16   | -      | -              |
| 1/16   | - s - - | 1/16   | -      | -              |
| 1/16   | - - s - | 1/16   | -      | -              |
| 1/16   | - - - t | -      | 1/16   | -              |
| 1/16   | - - - - | -      | -      | 1/16           |
| totals |         | 25/32  | 5/32   | 1/16           |

In 11 cases of these 16 cases, the more frequent outcome will be *s,* while in one case, *t* will be the more frequent. In three cases, we get a tie between *s* and *t,* so we split the probability in those cases. And in one case, we forget all four occurrences. In that case, we are unable to make a prediction. We represent this as the null outcome (Ø) in the list of possibilities: *s, t,* and Ø. Given an imperfect memory of $r = 1/2$, the overall probability that natural statistics predicts $s$ as the more frequent outcome therefore equals 25/32.

Natural statistics ends up making predictions that are equivalent to standard statistical decision theory, which sets up various levels of significance to represent the probability that a null hypothesis should not be rejected. In this particular problem, the null hypothesis (from the natural statistics point of view) states that the more frequent outcome $s$ is not more probable than the less frequent outcome $t$. There is more impreciseness in the natural statistics approach since there is a probability of predicting no outcome (in the above example, $p(\emptyset) = 1/16$). Asymptotically, the same predictions are made as in standard statistics, but only when the probability of remembering is one-half.

Once more the obvious question is why natural statistics should be equivalent to traditional statistics only when $r = 1/2$. This result naturally follows if each exemplar (or occurrence in the data) is accessed via a quantum bit (qubit) which is in either a spin-up (↑) or a spin-down (↓) state, and for which only one of these two states will permit accessibility. We suppose that each qubit has an equal chance of being in one of these two states. The direct asymptotic consequences will be that (1) the variance for estimating the probability of an outcome will be the standard unbiased estimate of variance, and (2) predicting the most frequent outcome will be the same as in standard statistical decision theory.

Accessibility to data also solves another difficult problem, that of randomness itself. In simulations of probabilistic behavior, computers use complicated pseudo-random functions to produce an apparently non-random sequence of integers. Such a sequence may appear random for long strings, but ultimately it is not random, but instead is fully predictable (by the pseudo-random function). It is wholly psychologically implausible that these complicated pseudo-random functions might be directly used by humans to predict non-deterministic language behavior.

On the other hand, true randomness is inherent at the quantum level. By providing random access to an occurrence (or to a pointer to an occurrence) in terms of qubits, we get actual randomness. In his descriptions of random selection as a rule of usage, Skousen never stated how the speaker would in fact be able to randomly select an occurrence (or a pointer to an occurrence). The problem of randomness was ignored in his initial work (Skousen 1989:37 and 1992:222). But by making an occurrence (or its pointer) accessible only when the assigned qubit is, say, in a spin-up state, actual randomness could be achieved. Furthermore, the statistical results would be asymptotically the same as standard statistics when we assume that the chances of the two qubit states, spin-up and spin-down, are equal.

**Probabilities in quantum mechanics, pointers in analogical modeling**

In analogical modeling, there is a lattice of supracontexts, partially ordered by the relationship of set inclusion. This lattice is defined by the given context, which is the set of variables for which we are trying to predict the outcome. Given $n$ variables in the given context, there are $2^n$ possible (unordered) combinations of those variables. In analogical modeling, each one of these possible combinations is called a supracontext. For instance, in attempting to predict the pronunciation of the initial $c$ of the word *ceiling* in terms of the 3 letters *eil* following the $c$ (namely, *eil*), we set up $2^3 = 8$ supracontexts for this given context (*eil*). For each supracontext we identify which exemplars belong and note their pronunciation of the initial $c$ letter, such as the /k/ sound for *coin,* the /s/ sound for *cell,* and the *ch* sound (represented as /č/) for *chin,* as follows:

|   |     | linear | | | squared | | | exemplars |
|---|-----|---|---|---|---|---|---|---|
|   |     | k-c | s-c | č-c | k-c | s-c | č-c |   |
|   | eil | -   | -   | -   |     |     |     |   |
|   | ei- | -   | -   | -   |     |     |     |   |
|   | e-l | -   | 1   | -   | 0   | 1   | 0   | *cell* |
|   | -il | -   | -   | -   |     |     |     |   |
|   | e-- | -   | 3   | -   | 0   | 9   | 0   | *cell, cent, certain* |
|   | -i- | 1   | -   | 1   | 2   | 0   | 2   | *chin, coin* |
| x | --l | 1   | 3   | -   |     |     |     |   |
| x | --- | 21  | 9   | 3   |     |     |     |   |

Some of these supracontexts have no occurrences (*eil, ei-,* and *-il*). Some have only one type of outcome (*e-l* and *e--*) and are therefore deterministic in behavior. One (*-i-*) is non-deterministic, yet has no subcontext that behaves differently. This kind of non-deterministic supracontext and the deterministic ones are homogeneous in behavior. Finally, there are some supracontexts (*--l* and *---*) for which there is at least one subcontext that behaves differently. Such non-deterministic supracontexts are heterogeneous. The *x*'s placed in front of the last two supracontexts mark these two supracontexts as heterogeneous.

In quantum computing, we will have *n* qubits for a given context of *n* variables, but these *n* qubits, unlike *n* classical bits, will allow us to simultaneously represent $2^n$ states -- namely, the superposition of all possible supracontexts. The advantage of quantum computing is that it allows massive simultaneity.

Each qubit has two states for each variable *i:*

| | | | |
|---|---|---|---|
| spin up | ↑ | 1 | variable *i* in supracontext |
| spin down | ↓ | 0 | variable *i* zeroed out |

These qubit variables defined by the given context are not assigned their spin-up and spin-down states independently of each other. Instead, there are important correlations between the qubits (referred to in quantum mechanics as entanglement). Moreover, each qubit is normally in a probabilistic state, a mixture of spin up and spin down.

For each of the $2^n$ supracontexts, we assign an amplitude. Ultimately, when we come to observe our lattice of supracontexts, we can require that the squares of these supracontextual amplitudes are normed; that is, the sum of the squared amplitudes equals one. This norming basically requires that for each supracontext the squared amplitude represents the probability of selecting that supracontext. The norming is really only necessary because probabilities themselves are mathematically defined as normed -- that is, as a measure on the line [0,1].

One important requirement for applying quantum computing to analogical modeling is that all empty and heterogeneous supracontexts must end up with zero amplitude (equivalent to zero probability of being selected). We need, of course, reversible operators to zero out heterogeneous supracontexts and make sure the empty supracontexts remained zeroed out. The remaining homogeneous supracontexts will, of course, show entanglement between the qubits representing the variables.

The first important connection between analogical modeling and quantum computing is that the number of occurrences assigned to a given supracontext is equivalent to the amplitude. In other words, linearity in analogical modeling corresponds to the amplitude in quantum computing. In our example for *ceiling*, we have the following amplitudes prior to norming, but after determining heterogeneity:

|   |     |       | *k-c* | *s-c* | *č-c* | amplitude ≡ occurrences | |
|---|-----|-------|-------|-------|-------|-----|-----|
|   | 111 | ↑↑↑   | -     | -     | -     | 0   | empty |
|   | 110 | ↑↑↓   | -     | -     | -     | 0   | empty |
|   | 101 | ↑↓↑   | -     | 1     | -     | 1   | deterministic |
|   | 011 | ↓↑↑   | -     | -     | -     | 0   | empty |
|   | 100 | ↑↓↓   | -     | 3     | -     | 3   | deterministic |
|   | 010 | ↓↑↓   | 1     | -     | 1     | 2   | non-deterministic |
| *x* | 001 | ↓↓↑   | 1     | 3     | -     | 0   | heterogeneous |
| *x* | 000 | ↓↓↓   | 21    | 9     | 3     | 0   | heterogeneous |

The requirement of normality means that the actual amplitude is the frequency of occurrence divided by the square root of the sum of the squared frequencies ($\sqrt{\Sigma x^2}$) -- in this case, the norming fraction is $1/\sqrt{14}$ since $1^2 + 3^2 + 2^2 = 14$.

Thus the number of occurrences in each homogeneous supracontext (the linear measure) is proportionally related to the amplitude. We can therefore give an alternative representation using Schrödinger's wave equation Ψ (in Dirac's notation). In the following example, each occurring homogeneous supracontext (101, 100, 010) is represented as a possible state:

$$|\Psi> = 1/\sqrt{14}\ |101> + 3/\sqrt{14}\ |100> + 2/\sqrt{14}\ |010>$$

Now in quantum computing, the probability of occurrence for each homogeneous supracontext will be the square of the amplitude. In order to predict the behavior of our system, we need to select a single supracontext from our superposition of $2^n$ supracontexts. In other words, observational decoherence of the superposition is equivalent to selecting an occurring homogeneous supracontext, but instead of using occurrences to make the selection, we use pointers to do that. In other words, the squaring of the amplitude in quantum computing is equivalent to selecting a pointer to an occurrence rather than selecting an occurrence directly. This means that if we use quantum computing to do analogical modeling, we will always be selecting the squaring function of analogical modeling. Earlier work in analogical modeling allowed either linearity or squaring (Skousen 1992:8-9), but now the choice of squaring over linearity is motivated.

In our example, decoherence of the superposition therefore leads to a probability. The probability of each supracontext is proportional to the square of the number of occurrences in that supracontext (in other words, proportional to the number of pointers to occurrences in that supracontext):

|   |     |       | k-c | s-c | č-c | probability ≡ pointers |                   |
|---|-----|-------|-----|-----|-----|-----|-------------------|
|   | 111 | ↑↑↑ | -   | -   | -   | 0   | empty             |
|   | 110 | ↑↑↓ | -   | -   | -   | 0   | empty             |
|   | 101 | ↑↓↑ | -   | 1   | -   | 1   | deterministic     |
|   | 011 | ↓↑↑ | -   | -   | -   | 0   | empty             |
|   | 100 | ↑↓↓ | -   | 9   | -   | 9   | deterministic     |
|   | 010 | ↓↑↓ | 2   | -   | 2   | 4   | non-deterministic |
| x | 001 | ↓↓↑ | 3   | 12  | -   | 0   | heterogeneous     |
| x | 000 | ↓↓↓ | 693 | 297 | 99  | 0   | heterogeneous     |

By norming the number of pointers, we get the following probabilistic predictions using quantum analogical modeling:

|  | probability | k-c | s-c | č-c | exemplars |
|---|---|---|---|---|---|
| 101 | $(1/\sqrt{14})^2 = 1/14$ | 0 | 1 | 0 | *cell* |
| 100 | $(3/\sqrt{14})^2 = 9/14$ | 0 | 9 | 0 | *cell, cent, certain* |
| 010 | $(2/\sqrt{14})^2 = 4/14$ | 2 | 0 | 2 | *chin, coin* |

The probabilities, of course, represent the squares of the amplitudes given by Schrödinger's wave equation $\Psi$:

$$|\Psi\rangle = 1/\sqrt{14}\ |101\rangle\ +\ 3/\sqrt{14}\ |100\rangle\ +\ 2/\sqrt{14}\ |010\rangle$$

Prediction in analogical modeling will also require each supracontext to be linked to actual exemplars.

**Measuring uncertainty in terms of disagreement**

The normal approach to measuring uncertainty has been to use Shannon's "information", which is equivalent to the entropy of classical statistical mechanics. This measure is a logarithmic measure (of the form $\Sigma\ p \log p$, where $p$ is the probability of a particular outcome). Shannon's uncertainty is equivalent to the number of yes-no questions needed (on the average) to determine the outcome. The natural interpretation of this measure is that one gets an unlimited number of chances to discover the correct outcome, an unreasonable possibility for a psychologically based theory of behavior. Furthermore, the entropy for continuous probabilistic distributions is always infinite. This last property forced Shannon to come up with an artificial definition for the entropy of a continuous distribution. (See the discussion in sections 1.11 and 3.8 of *Analogy and Structure*.)

In chapters 1-3 of *Analogy and Structure* (written in 1983, published in 1992), Skousen developed a quadratic measure of uncertainty called disagreement. This measure was applied to language behavior in *Analogical*

*Modeling of Language* (written in 1987 and published in 1989). The measure of disagreement is the probability of two randomly chosen occurrences disagreeing in outcome (namely, $1 - \Sigma\, p^2$, where once more $p$ is the probability of an outcome). There is a corresponding measure of agreement, namely the probability of agreement in outcome for two randomly chosen occurrences (that is, $\Sigma\, p^2$). The natural interpretation of these quadratic measures is that one gets a single chance to guess the correct outcome. Further, the agreement density for a continuous probabilistic distribution $f(x)$ is easily and naturally defined as $\int f^2(x)\, dx$. This measure of agreement almost always exists. In fact, it is a much better measure of variation for a continuous distribution than the traditional variance (Skousen 1992:83-84).

This same quadratic measure of agreement is found in Schrödinger's wave equation as $\int |\psi(x)|^2\, dx$. In order to get an overall probability of one, the integral over the entire space is normed, but still it is the squaring function that is used to determine the probability of a subspace. Analogical modeling uses this squaring function to measure the agreement density for a continuous probability distribution (see chapter 3 of *Analogy and Structure*). If Schrödinger's wave equation is a real function (rather than the more general case allowing complex functions), we get the same precise formulation for the density agreement found in *Analogy and Structure,* but without the norming (namely, $\int \psi^2(x)\, dx$).

**Reversible operators**

Having determined that there seems to be some extraordinarily close connections between analogical modeling and quantum computing, we turn to how we might define appropriate quantum operators for determining the analogical set of homogeneous supracontexts.

In designing a quantum computational system for analogical modeling, every operator meets the following two requirements:

(1) *simultaneity:* each operator must be defined so that it can apply simultaneously to each of the $2^n$ supracontexts;

(2) *reversibility:* each operator must be reversible.

The first requirement allows us to take advantage of the simultaneity of quantum computing. The second requirement basically means that no erasure of data is permitted prior to observation of the system (that is, prior to decoherence of the superpositioned supracontexts). Each data occurrence, after being read, must be kept. Any computational result must be recoverable, and by keeping all the input data, we insure recoverability.

Let us consider what we mean by a reversible operator. (The discussion in this section is for readers unfamiliar with quantum computing. The examples follow the explication in Berman, Doolen, Mainieri, and Tsifrinovich 1998:51-58.) The basic idea is that after an operator has applied, we are able to determine from the final (or output) state what initial (or input) state it came from. This requirement of recoverability basically means that there is a unique one-to-one connection between inputs and outputs, that no mergers or splits occur, only a shifting (or renaming, so to speak) of representations.

One clear example of a reversible operator is negation. An $n$-gate (where the $n$ stands for negation) is reversible because we simply switch or flip the polarity of a state $a$ (true to false and false to true). In the following listing, $a_i$ represents the initial state of $a$, while $a_f$ represents the final state of $a$:

$n$-gate

| $a_i$ | $a_f$ |
|---|---|
| 0 | 1 |
| 1 | 0 |

So given a final state $a_f$ of 0 (false), we know that $a_i$ was 1 (true); similarly, $a_f = 1$ implies that $a_i = 0$.

On the other hand, the *and*-operator is not reversible. With an *and*-gate, the final state $c_f$ is true (or 1) only if $a_i$ and $b_i$ were both true (or 1). If the final state is false (or 0), then there are three possible sets of initial states (00, 01, or 10), and we do not know which set of initial states produced the false output:

*and*-gate

| $a_i$ | $b_i$ | $c_f$ |
|---|---|---|
| 0 | 0 | 0 |
| 0 | 1 | 0 |
| 1 | 0 | 0 |
| 1 | 1 | 1 |

In quantum computing, however, we can construct a reversible gate that can be used as an *and*-gate. We do this by constructing what is called a *control-control-not* gate (or *ccn*-gate, for short). In this system, we switch the polarity of an initial state $c_i$ only if two other initial states $a_i$ and $b_i$ are each true. The initial states $a_i$ and $b_i$ act as control states and $c_i$ acts as a *not* state (thus, *control-control-not*). We get the following input-output relationships for the *ccn*-gate:

*ccn*-gate

| $a_i$ | $b_i$ | $c_i$ | $a_f$ | $b_f$ | $c_f$ |
|---|---|---|---|---|---|
| 0 | 0 | 0 | 0 | 0 | 0 |
| 0 | 0 | 1 | 0 | 0 | 1 |
| 0 | 1 | 0 | 0 | 1 | 0 |
| 0 | 1 | 1 | 0 | 1 | 1 |
| 1 | 0 | 0 | 1 | 0 | 0 |
| 1 | 0 | 1 | 1 | 0 | 1 |
| 1 | 1 | 0 | 1 | 1 | 1 |
| 1 | 1 | 1 | 1 | 1 | 0 |

For this reversible gate, there are eight possible sets of initial states and eight possible sets of final states. For the first six cases, the set of final states is identical to the set of input states (thus 000 → 000, 001 → 001, 010 → 010, 011 → 011, 100 → 100, 101 → 101). For the last two cases, we simply switch the polarity of the *c* state (thus 110 → 111 and 111 → 110). This results in a unique one-to-one function between all the sets of states. No information is lost, and from every set of output states we can determine the unique set of input states from which it was derived.

We also emphasize here that with a *ccn*-gate the two control states *a* and *b* make no change whatsoever. In a sense, these two states represent labels.

Now from this *ccn*-gate, we can define a reversible *and*-operator by considering only those cases where the initial state $c_i$ equals zero. Given the entire *ccn*-gate, we mark these four cases with a check mark:

|   | $a_i$ | $b_i$ | $c_i$ | $a_f$ | $b_f$ | $c_f$ |
|---|---|---|---|---|---|---|
| ✓ | 0 | 0 | 0 | 0 | 0 | 0 |
|   | 0 | 0 | 1 | 0 | 0 | 1 |
| ✓ | 0 | 1 | 0 | 0 | 1 | 0 |
|   | 0 | 1 | 1 | 0 | 1 | 1 |
| ✓ | 1 | 0 | 0 | 1 | 0 | 0 |
|   | 1 | 0 | 1 | 1 | 0 | 1 |
| ✓ | 1 | 1 | 0 | 1 | 1 | 1 |
|   | 1 | 1 | 1 | 1 | 1 | 0 |

If we isolate these four cases where $c_i = 0$, we can see that we have the equivalent of an *and*-gate:

*and*-gate (a *ccn*-gate with $c_i = 0$)

|   | $a_i$ | $b_i$ | $c_i$ | $a_f$ | $b_f$ | $c_f$ |
|---|---|---|---|---|---|---|
| ✓ | 0 | 0 | 0 | 0 | 0 | 0 |
| ✓ | 0 | 1 | 0 | 0 | 1 | 0 |
| ✓ | 1 | 0 | 0 | 1 | 0 | 0 |
| ✓ | 1 | 1 | 0 | 1 | 1 | 1 |

The basic difference between a non-reversible *and*-gate and a reversible *ccn*-gate acting as an *and*-gate is that in the reversible gate the input states *a* and *b* are carried over identically as output states. In other words, the initial information about the states *a* and *b* is kept intact in the reversible gate.

Reversibility essentially requires that we have to keep track of the input. Richard Feynman, one of the first who proposed applying quantum mechanics to computing, realized that reversibility meant that the input would be reproduced along with the output at the end of the computation:

> But note that input data must typically be carried forward to the output to allow for reversibility. Feynman showed that in general the amount of extra information that must be carried forward is just the input itself. (Richard Hughes, "Quantum Computation", in Hey 1999:196)

This result is of great significance for analogical modeling and, in fact, for all exemplar-based systems -- namely, reversibility leads to exemplar-based systems. If some form of quantum computing is used for language prediction, then all the exemplars used in a computation must be recoverable (at least up until decoherence). Quantum computation of any language-based system will therefore be an exemplar-based one, even if the system ends up acting as a neural net or as a set of rules.

**Quantum analogical modeling**

Within analogical modeling, a supracontext is heterogeneous whenever any subcontextual analysis of that supracontext leads to an increase in disagreement (Skousen 1989:23-37). It turns out that this decision procedure is equivalent to the most powerful test possible. However, by introducing imperfect memory (equal to one-half), the power of the test can be reduced to standard statistical testing (Skousen 1998:247-250).

This single decision procedure can be re-interpreted so that no mathematical calculation is ever involved; not even a measurement of disagreement between occurrences is necessary. This reworking of the procedure for determining homogeneity was discussed in both *Analogical Modeling of Language* and *Analogy and Structure* (Skousen 1989:33-35 and 1992:295-300). There it was shown that there are two types of homogeneous supracontexts: (1) the supracontext is deterministic in behavior (only one outcome occurs); (2) if the supracontext is non-deterministic, all its non-deterministic behavior is restricted to a single subcontext (or subspace). In the original algorithm for analogical modeling, testing for the second type of supracontext required the program to do a layered comparison between adjacent levels of supracontexts (that is, between supracontexts representing a difference of one variable). Such an algorithm guaranteed an exponential explosion in running time.

More seriously, from a quantum computing perspective, such an algorithm could never be re-interpreted in terms of reversible operators applying simultaneously to all the supracontexts at once. By shifting the perspective to quantum computing, Skousen was able to discover that by keeping track of only two factors for a supracontext (the first outcome and the first intersect, to be explained in the next section), homogeneity could be determined by reversible operators applying simultaneously. Moreover, the original algorithm initially assigned each occurrence to the supracontext closest to the given context. Within the quantum algorithm, each occurrence is simultaneously assigned to every supracontext that can possibly include the occurrence. This simultaneity avoids the "trickle-down" effect of the original layered algorithm, which also contributed to the exponential running time of the original approach.

**Quantum computing of analogical modeling**

We now see how the principles of quantum computing can be used to solve the exponential problems (in both memory and running time) for analogical modeling. This demonstration will be done in terms of the simple example from section 2.2 of *Analogical Modeling of Language* (Skousen 1989:23-37). The dataset there has five occurrences, each specified by three variables (composed of numbers the *0,1,2,3*) and an outcome, either *e* or *r:*

    **dataset**        310e       *m* (= 5)
                                  032r
                                  210r
                                  212r
                                  311r

We let *m* represent the number of accessed occurrences in the dataset. We will assume that these five occurrences were randomly selected from a larger database at a level of imperfect memory of one-half. This corresponds to the idea that data occurrences are accessed through, say, a spin-up state (given two equally probable quantum states, spin-up and spin-down, for database access).

We make predictions in terms of a given context. We let *n* stand for the number of variables found in the given context. In the example from *Analogical Modeling of Language,* the given context is *312:*

 **given context**   312   $n$ (= 3)

No outcome (*e* or *r*) is specified for the given context since that is what we are trying to predict. Our task then is to predict the outcome, either *e* or *r*, in terms of the three variables *312*. In this example, *n* is three.

For this given context, we now define $2^n$ (= $2^3$ = 8) supracontexts by means of *n* (= 3) qubits. The supracontexts specify a powerset -- namely, all the possible groupings of variables that can be theoretically used to predict the outcome for the given context. Initially, each of these *n* (= 3) qubits are equally assigned to two possible random states, one or zero. For a given supracontext, if a variable is assigned a one (1), this means that that variable is used to help define the contents of that supracontext. On the other hand, a zero (0) means that that variable will be completely ignored for that supracontext.

For the given context *312,* we therefore have the following $2^n$ (= $2^3$ = 8) supracontexts:

 **supracontexts**   111   $\exp(n) = 2^n$ (= $2^3$ = 8)
           110
           101
           011
           100
           010
           001
           000

The supracontext *110,* for example, means that the first two variables will be considered, but the third one will be ignored.

As we read each data occurrence, we determine its intersect with the given context. For instance, the first data occurrence is *310e*. When compared with the given context *312,* we see that the first two variables agree, but the last one does not. The corresponding intersect for *310e* and *312* is therefore *110.* For our five data occurrences (*310e, 032r, 210r, 212r,* and *311r*), we have the following five intersects:

| | | | |
|---|---|---|---|
| **intersects** | 110 | 310e & 312 | $m (= 5)$ |
| | 001 | 032r & 312 | |
| | 010 | 210r & 312 | |
| | 011 | 212r & 312 | |
| | 110 | 311r & 312 | |

For each data occurrence, we also record its outcome, whether it is *e* or *r*.

For each supracontext, we need to determine certain kinds of information, but only using reversible operators that can simultaneously apply to all of the $2^n$ supracontexts. In order to derive the analogical set (a superposition of all the possible supracontexts), we determine the following information as each data occurrence is read:

- include
- sum
- first outcome
- plurality of outcome
- first intersect
- plurality of intersect
- heterogeneity
- amplitude

In each case, we assign a qubit or a register of qubits to store this information for the supracontexts.

We first discuss how we determine the *include* qubit. As we read each data occurrence, we need to determine which of the $2^n (= 2^3 = 8)$ supracontexts the occurrence can be assigned to. We do this by defining a register of $n (= 3)$ qubits, which we refer to as the *contain* register. Initially, each qubit in this register is assigned a one, but when a data occurrence is read, some of these ones will be changed to zeros, depending on the intersect for that data occurrence. From this evolved *contain* register of qubits, we can then determine whether we include the occurrence in each of the supracontexts.

We start out then by determining which supracontexts the first data occurrence (*310e*) will be included in. We assign an individual *include* qubit to *310e,* initially set to ones:

|  | supracontext | contain | include |
|---|---|---|---|
| **data occurrence** |  |  | 310e |
| **intersect** | *110* |  | 110 |
|  | 111 | 111 | 1 |
|  | 110 | 111 | 1 |
|  | 101 | 111 | 1 |
|  | 011 | 111 | 1 |
|  | 100 | 111 | 1 |
|  | 010 | 111 | 1 |
|  | 001 | 111 | 1 |
|  | 000 | 111 | 1 |

As already noted, the intersect for *310e* is *110*. In our representation above, the intersect *110* is placed under the data occurrence *310e,* but for convenience' sake we also take the intersect of the data occurrence being currently considered and place it right above the eight supracontexts. The intersect *110* is used to determine which supracontexts will include the data occurrence *310e*. This occurrence *310e* should be contained in four supracontexts: *110, 100, 010,* and *000.* We determine which supracontexts contain this data occurrence by applying the following reversible operator for each of the $n$ (= 3) variables:

**contain**

    for $i = 1$ to $n$
        if *intersect* $[i] = 0$ and *supracontext* $[i] = 1$
        then *contain* $[i] = 0$

This operator means that when we compare *110* (the intersect for *310e*) with the eight supracontexts, we need only deal with the intersect variables that are zero (0). Now if the corresponding supracontextual variable is a one (1), we change the corresponding *contain* variable from its initial one (1) to zero (0). Reversibility is obtained because we do not change the actual supracontextual specifications, but instead use the *contain* register as "work space".

Applying this operator simultaneously to all eight supracontexts gives us the following evolution in the *contain* register:

|  | supracontext | contain | include |
|---|---|---|---|
| **data occurrence** |  |  | 310e |
| **intersect** | *110* |  | 110 |
|  | 111 | 111→110 | 1 |
|  | 110 | 111→111 | 1 |
|  | 101 | 111→110 | 1 |
|  | 011 | 111→110 | 1 |
|  | 100 | 111→111 | 1 |
|  | 010 | 111→111 | 1 |
|  | 001 | 111→110 | 1 |
|  | 000 | 111→111 | 1 |

Now if for a given supracontext the *contain* register has only ones, then the data occurrence *310e* will be contained within that supracontext. After applying the *contain* operator three times (once for each variable), we correctly get *111* for the supracontexts *110, 100, 010,* and *111*. The occurrence *310e* will therefore be included in these four supracontexts, but not the other four.

To determine the actual *include* qubit for a data occurrence, we need to use the *include* operator $n$ (= 3) times, once for each qubit in the *contain* variable (that is, once for each variable):

**include**

> for *i* = 1 to *n*
>     if *contain* [*i*] = 0
>     then *include* = 0

In the example of *310e,* of course, only the third qubit of the *contain* register has any zeros. So by using the *include* operator, the *include* qubit for *310e* becomes correctly set:

|  | supracontext | contain | include |
|---|---|---|---|
| **data occurrence** |  |  | 310e |
| **intersect** | *110* |  | 110 |
|  | 111 | 110 | 0 |
|  | 110 | 111 | 1 |
|  | 101 | 110 | 0 |
|  | 011 | 110 | 0 |
|  | 100 | 111 | 1 |
|  | 010 | 111 | 1 |
|  | 001 | 110 | 0 |
|  | 000 | 111 | 1 |

In order to continue using the *contain* register to determine the *include* qubit for the next data occurrence, we need to reset the *contain* register to all ones. We do this by reversely applying the *contain* operator:

**reverse contain**

> for *i* = 1 to *n*
>     if *intersect* [*i*] = 0 and *supracontext* [*i*] = 1
>     then *contain* [*i*] = 1

Applying this reversed operator, we get the original initial state for the *contain* register:

| | supracontext | contain | include |
|---|---|---|---|
| **data occurrence** | | | 310e |
| **intersect** | *110* | | 110 |
| | 111 | 110→111 | 0 |
| | 110 | 111→111 | 1 |
| | 101 | 110→111 | 0 |
| | 011 | 110→111 | 0 |
| | 100 | 111→111 | 1 |
| | 010 | 111→111 | 1 |
| | 001 | 110→111 | 0 |
| | 000 | 111→111 | 1 |

Before we read the next data occurrence, we determine the amplitude for each of the possible supracontexts at this stage of the quantum evolution. To do this, we set up a number of qubit registers that designate the following information for each supracontext: *the sum, the first outcome, the plurality of the outcomes, the first intersect, the plurality of the intersects, the heterogeneity,* and *the amplitude.* Initially, prior to considering any data occurrence, these qubit registers are all equally assigned zeros:

| supracontext | sum | outcome | | intersect | | hetero | ampl |
| | | 1st | plur | 1st | plur | | |
|---|---|---|---|---|---|---|---|
| 111 | 0 | - | 0 | - | 0 | 0 | 0 |
| 110 | 0 | - | 0 | - | 0 | 0 | 0 |
| 101 | 0 | - | 0 | - | 0 | 0 | 0 |
| 011 | 0 | - | 0 | - | 0 | 0 | 0 |
| 100 | 0 | - | 0 | - | 0 | 0 | 0 |
| 010 | 0 | - | 0 | - | 0 | 0 | 0 |
| 001 | 0 | - | 0 | - | 0 | 0 | 0 |
| 000 | 0 | - | 0 | - | 0 | 0 | 0 |

After we have determined which supracontexts include a particular data occurrence, we then apply the following operators simultaneously to each supracontext. In each case, we give the same name to the operator as the name of the qubit register that stores the result:

**sum**

    if *include* = 1 for the current data occurrence
    then increment *sum* by one
        // that is, *sum* = *sum* + 1 //

**first outcome and plurality of outcome**

    if *first outcome* is empty
    then store the outcome of the data occurrence
        in *first outcome*
    otherwise   // *first outcome* is filled //
    set *plurality of outcome* equal to one

**first intersect and plurality of intersect**

    if *first intersect* is empty
    then store the intersect of the data occurrence
        in *first intersect*
    otherwise   // *first intersect* is filled //
    set *plurality of intersect* equal to one

**heterogeneity**

    if both *plurality of outcome* and *plurality of intersect* equal one
    then set *heterogeneity* equal to one

**amplitude**

    if *heterogeneity* = 1
    then *amplitude* = 0
    otherwise   // *heterogeneity* = 0 //
    set *amplitude* equal to *sum*

Three of these registers can be represented by a single qubit (namely, *plurality of outcome, plurality of intersect,* and *heterogeneity*). The others need to contain specific qubit representations of various information:

| | |
|---|---|
| *sum* | zero or a positive integer |
| *first outcome* | an outcome |
| *first intersect* | an *n*-bit representation |
| *amplitude* | zero or a positive integer |

It should be noted that each of these could be accessed by a single qubit plus some associated informational register whenever the qubit is set to one:

*sum*  
    0 if there are no occurrences  
    1 if there is at least one occurrence  
        register gives *sum*

*first outcome*  
    0 if no (first) outcome has yet been found  
    1 if a first outcome has been found  
        register gives *first outcome*

*first intersect*  
    0 if no (first) intersect has yet been found  
    1 if a first intersect has been found  
        register gives *first intersect*

*amplitude*  
    0 if there is no amplitude  
    1 if there is an amplitude  
        register gives *amplitude*  
            (the same integer as in *sum*)

It should also be noted here that certain states, once reached for a given supracontext, are not changed throughout the evolution of the superpositioned system (up through decoherence or observation). Suppose we use the single-qubit system, as just described. Then whenever any of the following qubits is set to one, that qubit and any associated register will never be changed as long as the superposition is maintained: *first outcome, plurality of outcome, first intersect, plurality of intersect,* and *heterogeneity.* The value for *sum* for a given supracontext, on the other hand, will never decrease. The value for *amplitude* will also never decrease except when heterogeneity is achieved, in which case the amplitude will be immediately reduced to zero. And from then on, the amplitude for this supracontext will always remain at zero.

As long as we keep track of all the data occurrences in the dataset, all these operators are reversible. This reversibility basically means that our system must be an exemplar-based system of prediction if we are going to use quantum computing to determine the analogical set for a given context.

We now read, one at a time, the five data occurrences (*310e, 032r, 210r, 212r,* and *311r*). For each data occurrence, we first compare it with the given context *312* and determine the intersect for that occurrence, then apply the sequence of operators (*contain, include, sum, first outcome, plurality of outcome, first intersect, plurality of intersect, heterogeneity,* and *amplitude*) and finally at the end of the sequence reverse the *contain* register (that is, apply the operator *reverse contain*) before reading the next data occurrence.

**given context:** 312

<u>initial</u> state (no data occurrences read yet)

| **supracontext** | **contain** | **include** (non-read \| read) | | | | | **sum** |
|---|---|---|---|---|---|---|---|
| | | 311r | 212r | 210r | 032r | 310e | |
| 111 | 111 | 1 | 1 | 1 | 1 | 1 | 0 |
| 110 | 111 | 1 | 1 | 1 | 1 | 1 | 0 |
| 101 | 111 | 1 | 1 | 1 | 1 | 1 | 0 |
| 011 | 111 | 1 | 1 | 1 | 1 | 1 | 0 |
| 100 | 111 | 1 | 1 | 1 | 1 | 1 | 0 |
| 010 | 111 | 1 | 1 | 1 | 1 | 1 | 0 |
| 001 | 111 | 1 | 1 | 1 | 1 | 1 | 0 |
| 000 | 111 | 1 | 1 | 1 | 1 | 1 | 0 |

| | **outcome** | | **intersect** | | | |
|---|---|---|---|---|---|---|
| *supracontext* | **1st** | **plur** | **1st** | **plur** | **hetero** | **ampl** |
| 111 | - | 0 | - | 0 | 0 | 0 |
| 110 | - | 0 | - | 0 | 0 | 0 |
| 101 | - | 0 | - | 0 | 0 | 0 |
| 011 | - | 0 | - | 0 | 0 | 0 |
| 100 | - | 0 | - | 0 | 0 | 0 |
| 010 | - | 0 | - | 0 | 0 | 0 |
| 001 | - | 0 | - | 0 | 0 | 0 |
| 000 | - | 0 | - | 0 | 0 | 0 |

<u>intersect</u> (after 1st data occurrence read)
310 & 312 = 110

| **supracontext** | **contain** | **include** (non-read | read) | | | | **sum** | |
|---|---|---|---|---|---|---|---|---|
| | | 311r | 212r | 210r | 032r | 310e | | |
| *110* | | | | | | 110 | | |
| 111 | 111→110 | 1 | 1 | 1 | 1 | 0 | | 0 |
| 110 | 111→111 | 1 | 1 | 1 | 1 | 1 | | 1 |
| 101 | 111→110 | 1 | 1 | 1 | 1 | 0 | | 0 |
| 011 | 111→110 | 1 | 1 | 1 | 1 | 0 | | 0 |
| 100 | 111→111 | 1 | 1 | 1 | 1 | 1 | | 1 |
| 010 | 111→111 | 1 | 1 | 1 | 1 | 1 | | 1 |
| 001 | 111→110 | 1 | 1 | 1 | 1 | 0 | | 0 |
| 000 | 111→111 | 1 | 1 | 1 | 1 | 1 | | 1 |

| | **outcome** | | **intersect** | | | |
|---|---|---|---|---|---|---|
| *supracontext* | **1st** | **plur** | **1st** | **plur** | **hetero** | **ampl** |
| 111 | - | 0 | - | 0 | 0 | 0 |
| 110 | e | 0 | 110 | 0 | 0 | 1 |
| 101 | - | 0 | - | 0 | 0 | 0 |
| 011 | - | 0 | - | 0 | 0 | 0 |
| 100 | e | 0 | 110 | 0 | 0 | 1 |
| 010 | e | 0 | 110 | 0 | 0 | 1 |
| 001 | - | 0 | - | 0 | 0 | 0 |
| 000 | e | 0 | 110 | 0 | 0 | 1 |

**reverse contain**

<u>intersect</u> (after 2nd data occurrence read)
032 & 312 = 001

| **supracontext** | **contain** | **include** (non-read | read) | | | | **sum** |
|---|---|---|---|---|---|---|---|
| | | 311r | 212r | 210r | 032r | 310e | |
| *001* | | | | | 001 | 110 | |
| 111 | 111→001 | 1 | 1 | 1 | 0 | 0 | 0 |
| 110 | 111→001 | 1 | 1 | 1 | 0 | 1 | 1 |
| 101 | 111→011 | 1 | 1 | 1 | 0 | 0 | 0 |
| 011 | 111→101 | 1 | 1 | 1 | 0 | 0 | 0 |
| 100 | 111→011 | 1 | 1 | 1 | 0 | 1 | 1 |
| 010 | 111→101 | 1 | 1 | 1 | 0 | 1 | 1 |
| 001 | 111→111 | 1 | 1 | 1 | 1 | 0 | 1 |
| 000 | 111→111 | 1 | 1 | 1 | 1 | 1 | 2 |

| | **outcome** | | **intersect** | | | |
|---|---|---|---|---|---|---|
| *supracontext* | **1st** | **plur** | **1st** | **plur** | **hetero** | **ampl** |
| 111 | - | 0 | - | 0 | 0 | 0 |
| 110 | e | 0 | 110 | 0 | 0 | 1 |
| 101 | - | 0 | - | 0 | 0 | 0 |
| 011 | - | 0 | - | 0 | 0 | 0 |
| 100 | e | 0 | 110 | 0 | 0 | 1 |
| 010 | e | 0 | 110 | 0 | 0 | 1 |
| 001 | r | 0 | 001 | 0 | 0 | 1 |
| 000 | e | 1 | 110 | 1 | 1 | 0 |

**reverse contain**

<u>intersect</u> (after 3rd data occurrence read)
210 & 312 = 010

| **supracontext** | **contain** | **include** (non-read | read) | | | | | **sum** |
|---|---|---|---|---|---|---|---|---|
| | | 311r | 212r | 210r | 032r | 310e | | |
| *010* | | | | 010 | 001 | 110 | | |
| 111 | 111→010 | 1 | 1 | 0 | 0 | 0 | | 0 |
| 110 | 111→011 | 1 | 1 | 0 | 0 | 1 | | 1 |
| 101 | 111→010 | 1 | 1 | 0 | 0 | 0 | | 0 |
| 011 | 111→110 | 1 | 1 | 0 | 0 | 0 | | 0 |
| 100 | 111→011 | 1 | 1 | 0 | 0 | 1 | | 1 |
| 010 | 111→111 | 1 | 1 | 1 | 0 | 1 | | 2 |
| 001 | 111→110 | 1 | 1 | 0 | 1 | 0 | | 1 |
| 000 | 111→111 | 1 | 1 | 1 | 1 | 1 | | 3 |

| | **outcome** | | **intersect** | | | |
|---|---|---|---|---|---|---|
| *supracontext* | **1st** | **plur** | **1st** | **plur** | **hetero** | **ampl** |
| 111 | - | 0 | - | 0 | 0 | 0 |
| 110 | e | 0 | 110 | 0 | 0 | 1 |
| 101 | - | 0 | - | 0 | 0 | 0 |
| 011 | - | 0 | - | 0 | 0 | 0 |
| 100 | e | 0 | 110 | 0 | 0 | 1 |
| 010 | e | 1 | 110 | 1 | 1 | 0 |
| 001 | r | 0 | 001 | 0 | 0 | 1 |
| 000 | e | 1 | 110 | 1 | 1 | 0 |

**reverse contain**

<u>intersect</u> (after 4th data occurrence read)
212 & 312 = 011

| **supracontext** | **contain** | **include** (non-read | read) | | | | **sum** |
|---|---|---|---|---|---|---|---|
| | | 311r | 212r | 210r | 032r | 310e | |
| *011* | | | 011 | 010 | 001 | 110 | |
| 111 | 111→011 | 1 | 0 | 0 | 0 | 0 | 0 |
| 110 | 111→011 | 1 | 0 | 0 | 0 | 1 | 1 |
| 101 | 111→011 | 1 | 0 | 0 | 0 | 0 | 0 |
| 011 | 111→111 | 1 | 1 | 0 | 0 | 0 | 1 |
| 100 | 111→011 | 1 | 0 | 0 | 0 | 1 | 1 |
| 010 | 111→111 | 1 | 1 | 1 | 0 | 1 | 3 |
| 001 | 111→111 | 1 | 1 | 0 | 1 | 0 | 2 |
| 000 | 111→111 | 1 | 1 | 1 | 1 | 1 | 4 |

| | **outcome** | | **intersect** | | | |
|---|---|---|---|---|---|---|
| *supracontext* | **1st** | **plur** | **1st** | **plur** | **hetero** | **ampl** |
| 111 | - | 0 | - | 0 | 0 | 0 |
| 110 | e | 0 | 110 | 0 | 0 | 1 |
| 101 | - | 0 | - | 0 | 0 | 0 |
| 011 | r | 0 | 011 | 0 | 0 | 1 |
| 100 | e | 0 | 110 | 0 | 0 | 1 |
| 010 | e | 1 | 110 | 1 | 1 | 0 |
| 001 | r | 0 | 001 | 1 | 0 | 2 |
| 000 | e | 1 | 110 | 1 | 1 | 0 |

**reverse contain**

```
intersect (after 5th data occurrence read)
311 & 312 = 110
```

| supracontext | contain | include (non-read \| read) | | | | | sum |
|---|---|---|---|---|---|---|---|
| | | 311r | 212r | 210r | 032r | 310e | |
| *110* | | 110 | 011 | 010 | 001 | 110 | |
| 111 | 111→110 | 0 | 0 | 0 | 0 | 0 | 0 |
| 110 | 111→111 | 1 | 0 | 0 | 0 | 1 | 2 |
| 101 | 111→110 | 0 | 0 | 0 | 0 | 0 | 0 |
| 011 | 111→110 | 0 | 1 | 0 | 0 | 0 | 1 |
| 100 | 111→111 | 1 | 0 | 0 | 0 | 1 | 2 |
| 010 | 111→111 | 1 | 1 | 1 | 0 | 1 | 4 |
| 001 | 111→110 | 0 | 1 | 0 | 1 | 0 | 2 |
| 000 | 111→111 | 1 | 1 | 1 | 1 | 1 | 5 |

| | outcome | | intersect | | | |
|---|---|---|---|---|---|---|
| *supracontext* | 1st | plur | 1st | plur | hetero | ampl |
| 111 | - | 0 | - | 0 | 0 | 0 |
| 110 | e | 1 | 110 | 0 | 0 | 2 |
| 101 | - | 0 | - | 0 | 0 | 0 |
| 011 | r | 0 | 011 | 0 | 0 | 1 |
| 100 | e | 1 | 110 | 0 | 0 | 2 |
| 010 | e | 1 | 110 | 1 | 1 | 0 |
| 001 | r | 0 | 001 | 1 | 0 | 2 |
| 000 | e | 1 | 110 | 1 | 1 | 0 |

**reverse contain**

In quantum mechanics the values of the amplitudes are systematically adjusted so that their squares sum to one. But as already pointed out, such norming procedures are the result of specifying that probabilities are real numbers from 0 to 1. In analogical modeling, there are no underlying probabilities, only occurrences and pointers to occurrences. Under conditions of imperfect memory, analogical modeling does produce probabilistic behavior, but without directly learning probabilities or using them. The amplitude for a homogeneous supracontext is directly proportional to the number of occurrences for that supracontext. Its probability of being selected is directly proportional to the number of pointers to occurrences in that supracontext -- which is the square of the number of occurrences.

This squaring occurs in quantum computing whenever decoherence occurs. But in quantum analogical modeling, the squaring does not involve mathematical calculation. Instead, it is the result of selecting from all the homogeneous supracontexts one of the pointers to occurrences. We do not select an occurrence itself; that kind of selection would lead to setting the probability of predicted outcome as proportional to the amplitude. Instead, we select a pointer to an occurrence, which gives the probability of predicted outcome as proportional to the amplitude squared.

Since the supracontexts are the quantum states, decoherence is equivalent to observing one of the homogeneous supracontexts, then selecting one of the pointers to occurrences in that supracontext. Nonetheless, it is worth noting that one could directly select one of the pointers to occurrences in any of any of the homogeneous supracontexts and get the same results -- namely, the proportional probability of random selection defined by the frequency squared. Furthermore, since we are keeping track of all the data occurrences, we do not really need to keep track of the sum and amplitude per se, only the information that determines the heterogeneity of each supracontext.

This view of decoherence rejects Shannon's unbounded measure of uncertainty, which allows an unlimited number of yes-no questions to guess the correct outcome. Analogical modeling allows only one guess and is equivalent to a measure of simple disagreement between pairs of occurrences. Analogical modeling thus looks at behavior in terms of events and connections between events (that is, as data occurrences and pointers between those occurrences). In analogical modeling, this measure of disagreement thus shows up directly whenever observation or decoherence occurs.

In our example, after reading the five data occurrences, we have two (non-occurring) supracontexts with no occurrences (where the sum equals zero), two heterogeneous supracontexts, and four occurring (non-zero) homogeneous supracontexts. When observation takes place, we randomly select one of these four non-zero homogeneous supracontexts in proportion to their number of pointers to occurrences:

```
      include                              sum    hetero   ampl   prob

      311r  212r  210r  032r  310e
      110   011   010   001   110

111   0     0     0     0     0    0      0        0      0
110   1     0     0     0     1    2      0        2      4
101   0     0     0     0     0    0      0        0      0
011   0     1     0     0     0    1      0        1      1
100   1     0     0     0     1    2      0        2      4
010   1     1     1     0     1    4      1        0      0
001   0     1     0     1     0    2      0        2      4
000   1     1     1     1     1    5      1        0      0
```

For each of the occurring homogeneous supracontexts, we can readily determine how many pointers point to each of the two possible outcomes, *e* and *r:*

```
        include                              ampl     prob     pointers

        311r   212r   210r   032r   310e                       e    r
        110    011    010    001    110

110     1      0      0      0      1        2        4        2    2
011     0      1      0      0      0        1        1        0    1
100     1      0      0      0      1        2        4        2    2
001     0      1      0      1      0        2        4        0    4
```

Thus the chances of selecting the outcome *e* (that is, the chances of selecting a pointer to an occurrence having the *e* outcome) is 4 (= 2+2), while the chances of selecting the outcome *r* (that is, the chances of selecting a pointer to an occurrence having the *r* outcome) is 9 (= 2+1+2+4). The probability of the *e* outcome is therefore 4/13 (≈ 0.31), and the probability of the *r* outcome is 9/13 (≈ 0.69). These are the same results derived in section 2.2 of *Analogical Modeling of Language* (Skousen 1989:23-37). The approach there, however, is based directly on the principle of minimizing the quadratic measure of disagreement. The quantum computational approach considers the plurality of outcomes and intersects, but derives the very same analogical set.

      We can see from the superposition of $2^n$ supracontexts that the exponential explosion of analogical modeling is reduced to a polynomial function of *n*. For each of *m* data occurrences accessed from the database, we will need a single *include* qubit. In terms of memory requirements, quantum analogical modeling will require a linear qubit size of O(*m*+*n*). On the other hand, the required running time is be O(*m·n*), a multiplicative function. However, for a set number of data occurrences, the running time will be a linear function of *n*, the number of variables. These results provide the tractability we need for a viable exemplar-based approach to language prediction.

      David Eddington, in a preface to his paper "Analogy and the Dual-Route Model of Morphology" given at the Conference on Analogical Modeling (and published as Eddington 2000a), has compared analogical modeling and its problem with exponentiality to a heavyweight boxer, very slow but powerful. However, if quantum computing can be applied to analogical modeling, we may have a heavyweight that is exponentially faster than anyone conceived of. Up to this time, we have perhaps worried too much about the exponential explosion, as if

this were a problem that must be solved by any other means. Quantum computing suggests that we treat the exponentiality of analogical modeling as inherent. Instead of trying to avoid the exponential explosion, we should embrace it!

**Analogical quantum mechanics**

It is also worth noting that analogical modeling may provide an interpretative model for quantum mechanics itself. As many have noted, the problem with quantum mechanics is that it is a formalism in search of an interpretation (see Cushing 1998:271-355, especially chapter 23).

Analogical modeling does not actually posit underlying probabilities -- there are no inherent probabilities. Instead, analogical modeling proposes occurrences (or events) and pointers (or connections) between occurrences. The notion of agreement and disagreement between occurrences leads to a natural measure of (un)certainty, one that directly models the linear/squared relationship of amplitudes and probabilities in quantum mechanics. The superpositioned supracontexts in analogical modeling, however, keep track of occurrences, not amplitudes per se. Decoherence leads to selecting a pointer to an occurrence. The probabilities are the result of selecting a pointer to an occurrence. Furthermore, the predictions are based on a single observation, as is analogical modeling (especially given its measure of disagreement instead of Shannon's information, which permits any number of observations). The norming of probabilities is not inherent to quantum mechanics. The real question is whether the results are probabilistic. Setting a norm on the probability measure is merely a mathematical convention. If one wishes, one can continually norm the amplitudes to obtain an observed probability between zero and one.

**Note**

I wish to thank members of the Analogical Modeling Research Group for their comments on the ideas of this paper: Dil Parkinson, Deryle Lonsdale, Theron Stanford, and Don Chapman.

# References

Aha, David W., Dennis Kibler, and Marc K. Albert (1991), Instance-Based Learning Algorithms. *Machine Learning* 6, 37-66.

Berman, Gennady P., Gary D. Doolen, Ronnie Mainieri, and Vladimir I. Tsifrinovich, eds. (1998), *Introduction to Quantum Computers*. Singapore: World Scientific.

Cushing, James T. (1998), *Philosophical Concepts in Physics: The Historical Relation between Philosophy and Scientific Theories*. Cambridge: Cambridge University Press.

Daelemans, Walter, Steven Gillis, and Gert Durieux (1994), The Acquisition of Stress: A Data-Oriented Approach. *Computational Linguistics* 20(3), 421-451.

Daelemans, Walter, Steven Gillis, and Gert Durieux (1997), Skousen's Analogical Modeling Algorithm: A Comparison with Lazy Learning. *New Methods in Language Processing* (eds. Daniel Jones and Harold Somers), 3-15. London: University College Press.

Derwing, Bruce, and Royal Skousen (1994), Productivity and the English Past Tense: Testing Skousen's Analogy Model. *The Reality of Linguistic Rules* (eds. Susan D. Lima, Roberta L. Corrigan, and Gregory K. Iverson), 193-218. Amsterdam: John Benjamins.

Eddington, David (2000a), Analogy and the Dual-Route Model of Morphology. *Lingua* 110, 281-298.

Eddington, David (2000b), Spanish Stress Assignment within Analogical Modeling of Language. *Language* 76, 92-109.

Hey, Anthony J. G. Hey, ed. (1999), *Feynman and Computation: Exploring the Limits of Computers*. Reading, Massachusetts: Perseus Books.

Jones, Daniel (1996), *Analogical Natural Language Processing*. London: University College London Press.

Lo, Hoi-Kwong, Sandu Popescu, and Tom Spiller, eds. (1998), *Introduction to Quantum Computation and Information*. Singapore: World Scientific.

McClelland, James L., David E. Rumelhart, and the PDP Research Group (1986), *Parallel Distributed Processing* [PDP], 2 volumes. Cambridge, Massachusetts: MIT Press.

Rytting, C. Anton (2000), An Empirical Test of Analogical Modeling: The /k/~Ø Alternation. *LACUS Forum* 26, 73-84.

Skousen, Royal (1989), *Analogical Modeling of Language*. Dordrecht: Kluwer.

Skousen, Royal (1992), *Analogy and Structure*. Dordrecht: Kluwer.

Skousen, Royal (1998), Natural Statistics in Language Modeling. *Journal of Quantitative Linguistics* 5, 246-255.

Skousen, Royal (in press), Analogical Modeling. *Quantitative Linguistics: An International Handbook* (eds. Gabriel Altmann, Reinhard Köhler, and Raimund G. Piotrowski). Berlin: Walter de Gruyter. [article currently available at <http://humanities.byu.edu/aml/homepage.html>]

Williams, Colin P. and Scott H. Clearwater (1998), *Explorations in Quantum Computing*. New York: Springer-Verlag.

Wulf, Doug (1996), An Analogical Approach to Plural Formation in German. *Proceedings of the Twelfth Northwest Linguistics Conference. Working Papers in Linguistics* 14, 239-254. Seattle: University of Washington.